\documentclass{article}
\usepackage{fleqn,espcrc2,epsfig}
\newcommand{\prb}{Phys. Rev. B}

\title{Search for effective models of stripes in the cuprates}

\author{Oleg Tchernyshyov        
	and 
	Leonid P. Pryadko\address{Institute for Advanced Study,
		School of Natural Sciences, \\
		Olden Lane, Princeton, NJ 08540, USA.}\thanks{
Research supported in part by DOE Grant DE-FG02-90ER40542.}
}
       
\begin{document}
\markright{\tt IASSNS-HEP 99/113} 
\thispagestyle{myheadings}

\begin{abstract}
We argue that effective 1D models of stripes in the cuprate
superconductors can be constructed by studying ground states and
elementary excitations of domain walls in 2D model antiferromagnets.
This method, applied to the $t$--$J$ model with Ising anisotropy,
yields two such limiting cases: an ordinary 1D electron gas and a 1D
gas of holons strongly coupled to transversal fluctuations of the
stripe.
\end{abstract}

\maketitle

Formation of charge stripes in some of the cuprate superconductors
\cite{Tranquada} is a peculiar phenomenon in its own right.  A
possible connection between the stripes and high-temperature
superconductivity makes them even more attractive to a theorist.  The
puzzle is nevertheless quite hard: to date there is no microscopic
theory describing the physics of stripes in the cuprates.

Some progress has been made.  Hartree-Fock studies of the Hubbard
model near half-filling \cite{early-theory} have revealed that doped
charges segregate in the form of narrow stripes.  The
stripes are domain walls separating antiferromagnetic (AF) domains
with opposite orientations of the Neel vector, in agreement with
experiments.  However, the predicted linear density of charge on a
stripe, $\nu=1$ hole per unit cell, indicates that there are no
charges able to carry current.  Experimentally, $\nu\approx 1/2$ for
non-overlapping stripes \cite{Yamada}.

In the absence of a microscopic theory, it is reasonable to look for a
model description in which a stripe is a 1D object
interacting with the surrounding antiferromagnet.  A stripe is
characterized by a ground state, by its elementary excitations, and by
the interactions of the excitations among themselves and with the
environment.

\section{The strategy}

A stripe is a complicated object.  Its degrees of freedom may include
charge and spin of the stripe particles, as well as transverse
fluctuations of its position.  

One potentially promising route to finding the right 1D model is to
study simple solvable models of an antiferromagnet in 2D in the
presence of a domain wall.  A single domain wall can be created
artificially, e.g., by wrapping a system with an odd number of rows on
a cylinder.  One may then study an isolated stripe at any filling
$\nu$ by adding or removing the right number of electrons.

A 2D model may look unrealistic, but with universality and a bit of
luck the resulting 1D theory may have just the right symmetry of the
vacuum and the correct quantum numbers of elementary excitations.

\section{An example}

Consider a 2D model, a variant of the $t$--$J$ model in the Ising
limit $J_z\gg J_\perp, t$ \cite{tJz}:
\begin{eqnarray}
  \label{t-J-Ham}
  H = \sum_{\langle{\bf rr'}\rangle}\Bigl[ 
    -t\, a^\dagger_{\sigma}({\bf r'}) \, a_{\sigma}({\bf r})
    + {J_\perp\over2}\, s_+({\bf r'})\, s_-({\bf r}) 
     \nonumber \\ 
   + \mbox{ H.\,c.} + \,  J_z\, s_z({\bf r'})\, s_z({\bf r}) 
  + V\, n({\bf r'}) \, n ({\bf r})\Bigr],
\end{eqnarray} 
with the usual exclusion of doubly occupied sites.  The dominating
$J_z$ term minimizes the number of frustrated (ferromagnetic) bonds
and thus sets the ground state.  The energy of an AF
bond $V-J_z/4$ controls interactions of charged quasiparticles
on the domain wall and in the bulk.  Making $V-J_z/4 \gg t>0$ 
prevents phase separation in the  bulk, as well as on a stripe.

The physics of a stripe in this model is very different in the
limits $\nu\to 1$ and $\nu\to 0$.  

\subsection{$\nu\to 1$: 1D electron gas}

Assuming a domain wall of length $L$ has been created by frustrating
boundary conditions, we first remove $L$ electrons.  The holes end up
on the wall forming a zigzag stripe with $\nu=1$.

\begin{figure}[htb]
\epsfxsize=\columnwidth
\epsffile{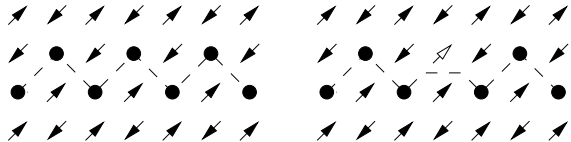}
\caption{A zigzag stripe in the $t$--$J_z$ model.  Left: $\nu=1$.
Right: $\nu=1-\varepsilon$.  Solid circles are sites with electrons
removed.  The dashed line is the location of the domain wall.  The
open arrow denotes an electron quasiparticle.}
\label{fig:electron}
\end{figure}

For $\nu = 1-\varepsilon$, the stripe will contain $\varepsilon L \ll
L$ quasiparticles with the quantum numbers of electrons, $Q=-1$, $S_3
= \pm 1/2$ (Fig.~\ref{fig:electron}).  Because of the zigzag geometry
of the stripe, the electron quasiparticles feel a strong (of order
$J_z$) staggered magnetic field.  It confines electrons with opposite
spins to different rows of the stripe.  Within its row, an electron
can delocalize to reduce kinetic energy.  In the tight-binding model
(\ref{t-J-Ham}), such hopping is a multistage process, which involves
pushing a neighbor spin out of the way.

Despite a suppressed hopping amplitude, the system can be described a
dilute 1D electron gas.

\markright{}

\subsection{$\nu\to 0$: 1D holon gas}

\begin{figure}[htb]
\epsfxsize=\columnwidth
\epsffile{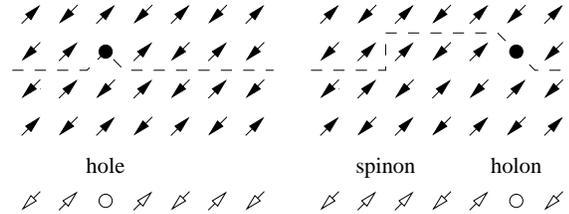}
\caption{A bond-centered stripe near $\nu=0$ in the $t$--$J_z$ model.
A doped hole (left) splits into a spinon and a mobile holon (right).
Open symbols at the bottom denote spin and charge integrated {\em
across} the domain wall with a smooth envelope function.  }
\label{fig:holon}
\end{figure}

Precisely at half-filling ($\nu=0$), the domain wall is bond-centered.
A single doped hole ends up at the domain wall (Fig.~\ref{fig:holon},
left).  If it starts to move alng the wall, two remarkable things
happen.  First, the hole leaves its spin $S_3=\pm1/2$ behind in the
form of a spinon and then propagates freely as a spinless object (a
holon, Fig.~\ref{fig:holon}, right).  No additional frustrated bonds
are produced afterwards.  Furthermore, no costly spinons would be left
behind had we started with 2 holes.  Second, the holon (as well as the
spinon) resides on a transverse kink of the domain wall.  These
elementary excitations are thus maximally strongly coupled to
transverse fluctuations of the stripe.  Kink direction can
be specified by assigning a holon (or a spinon) a transverse flavor
$\pm 1/2$ \cite{isospin}.

Holons are, in fact, the elementary excitations of the model, at
least at low doping $\nu\ll 1$: their kinetic energy is lower than
that of immobile holes.  

Finally, we note that holons with transverse flavor have also been
found at small $\nu$ in a Hartree-Fock study of the Hubbard model
\cite{9907472}.

\end{document}